# InAs/GaAs (211)B quantum dots with negligible FSS for the generation of entangled photons

S. Germanis[1], A. Beveratos[2], G.E. Dialynas[3], G. Deligeorgis[4], P.G. Savvidis[1,4], Z. Hatzopoulos[3,4], and N.T. Pelekanos[1,4]

[1]Department of Materials Science and Technology, University of Crete, P.O. Box 2208, 71003 Heraklion, Greece

[2]Laboratoire de Photonique et Nanostructures LPN-CNRS, UPR-20 Route de Nozay, 91460 Marcoussis, France

[3]Department of Physics, University of Crete, P.O. Box 2208, 71003 Heraklion, Greece

[4]Microelectronics Research Group, IESL-FORTH, P.O. Box 1385, 71110 Heraklion, Greece

Abstract

Polarization-resolved single dot spectroscopy performed on (211)B InAs/GaAs quantum dots reveals that the fine structure splitting of the excitonic levels in these dots is much lower compared to the usual (100)-grown InAs dots. Time-resolved measurements confirm the high oscillator strength of these dots, and thus their good quantum efficiency at 4 K, comparable with that of (100) InAs/GaAs dots. Last, photon correlation measurements demonstrate single photon emission out of the excitonic optical transition of these dots. All these features make this novel dot system very promising for implementing solid-state entangled photon sources.

Quantum information science strongly relies on the efficient generation of "on demand" single and entangled photon sources [1,2]. In several pioneering experiments [3–5], entangled photon pairs were obtained using parametric down-conversion [6,7], which, while easy to implement suffers from the Poissonian statistics of the emitted photon pairs, leading to multi-pair emission and thus decreasing the fidelity of entanglement [8]. On demand entangled photon pairs can be obtained by heralding a 3-photon process [3] or with the detection of several auxiliary photons [9], but both schemes are inefficient in terms of photon pair production probability. Single quantum dots (QDs) have been proposed as sources of "on demand" polarization entangled photon sources [10] by using the radiative decay of two electron-hole pairs trapped in the QD. The subsequent decay of the biexciton and exciton states will produce polarization entangled photon pairs as long as the two possible decay paths to the ground state $|\sigma^+_{XX}\rangle|\sigma^-_X\rangle$ and $|\sigma^-_{XX}\rangle|\sigma^+_X\rangle$ are indistinguishable. In actual QDs, due to the asymmetry of the dot, a fine structure splitting (FSS) of the intermediate exciton level arises, rendering the two paths distinguishable and diminishing the entangled photon pair fidelity [11,12]. Several techniques, using QD annealing [13], in-plane magnetic [14] or electric field [15] as well as perpendicular electric field [16] have been proposed in order to cancel the FSS. Purcell effect on the exciton line can also be used to broaden the exciton level and make the two paths indistinguishable [12]. Generation of polarization entangled photons has been reported in the last few years by using one or several of the above techniques [17-19]. While these techniques are successful, they are not scalable, and it is not simple to find a QD with FSS below 2μeV necessary for the generation of high-fidelity entangled photons [12].

Alternatively, it has been proposed recently that (111)-grown nanostructures are capable of negligibly small FSS values, due to the absence of symmetry-lowering effects in this orientation [20, 21]. Major role in this plays the fact that QDs grown along polar orientations, such as (111) or (211), contain very large piezoelectric (PZ) fields along the growth direction, of the order of MV/cm [22-24]. A recent preliminary report on (111) InGaAs/GaAs QDs fabricated using droplet epitaxy has shown evidence for reduced FSS values (10-40μeV) [25]. However, the multiple excitonic lines present in the single dot spectra as well as the lack of structural information about the emitting dots makes difficult the interpretation of these results. In this letter, we present results on InAs/GaAs QDs grown along the (211)B orientation in the Stranski-Krastanow growth mode, which exhibit FSS values smaller than our measuring resolution of

10μeV, *for over 90% of the dots*, suggesting that (211)B-grown QDs are suitable for use in entangled photon sources.

The sample used in this study is grown by molecular beam epitaxy on a (211)B semi-insulating GaAs substrate. It contains a single layer of InAs QDs, grown in the middle of a 10nm-thick GaAs/Al$_{0.3}$Ga$_{0.7}$As quantum well, about 50nm beneath the sample surface. The QD layer is grown by depositing 1.5 monolayers (MLs) of InAs at 500ºC with a growth rate of 0.1ML/s. Under these conditions, the InAs dots take the shape of truncated pyramids, with typical QD heights between 2 and 3nm, a diameter to height aspect ratio of about 10, and a QD density of ~10$^{10}$cm$^{-2}$ [26]. As an example, Fig.1a presents an atomic force microscopy (AFM) image from an uncapped QD sample, grown at precisely the same conditions, the analysis of which reproduces well the above values. For single dot spectroscopy, the sample is processed by e-beam lithography into square or circular mesas of various sizes (150-500nm), an example of which is shown in Fig.1b. A continuous-wave laser diode emitting at 405nm is used for the excitation of isolated nano-mesas with a spot size of ~5μm. The micro-photoluminescence (μ-PL) signal is dispersed in a 0.75m spectrograph with 1200gr/mm grating and is detected by a Nitrogen-cooled charge-coupled device (CCD) camera. The resolving power of this setup using standard de-convolution procedure is determined to be ~10μeV. The sample is cooled down to 8K in a variable temperature continuous-flow helium cryostat. The polarization-resolved spectra are recorded using a fixed linear polarizer in front of the spectrograph and a broadband λ/2 wave-plate to rotate the polarization. For the time resolved PL experiments, a mode-locked Ti:sapphire laser at 780nm was used with 4ps pulse-width and 81MHz repetition rate.

A characteristic μ-PL spectrum from a single (211)B InAs/GaAs QD is presented in Fig.1c. It consists of two main lines, labeled as X and XX, which have been identified as exciton and bi-exciton emission peaks, based on their linear and quadratic power dependence, respectively, at low powers. In fact, the intensity of X scales at low powers as ~P$^{0.89}$, whereas that of XX as ~P$^{1.8}$. The energy difference between the XX and X lines is ~4meV in this QD, with the XX line appearing characteristically at higher energy. This is a clear manifestation of the PZ field inside these dots, which has been measured by us to be as strong as ~1MV/cm [23, 24]. Considering that the (211) surface resembles closely to a (111) surface, we argue that the predictions of negligible FSS in (111)-grown QDs should equally apply to the (211) QD system.

To determine FSS in (211)B InAs/GaAs QDs, we performed polarization dependent μ-PL on 20 different single dots of the sample. Among these, only two have shown measurable FSS, i.e. larger than our resolving power of 10μeV. One such case is shown in the top of Fig.2, where the X and XX emission lines of Fig.1c are analyzed in terms of linear polarization. We observe that while the intensity of the lines remains practically intact, their energy positions vary periodically with polarizer angle, in an "out-of-phase" fashion. In the bottom of Fig.2, the XX-X energy difference is plotted as a function of polarizer angle, exhibiting the characteristic periodicity of 180°. From the max-min energy difference, we deduce that the FSS in this QD is 20μeV. However, as previously mentioned, 90% of the dots tested in this work have shown no measurable FSS. A typical example is depicted in Fig.3, where the X and XX emission lines of another dot appear completely insensitive to polarization. Since only 10% of the QDs have a measurable splitting of more than 10μeV, we argue that it is highly probable that the QDs have a splitting of less than 8μeV, assuming a gaussian distribution. This is to be compared to the mean FSS of 40μeV for (100) InAs/GaAs QDs [13], which can be reduced to less than 10μeV when annealing is applied to specific dots [27].

Negligible FSS is not a sufficient condition to make the (211)B InAs/GaAs QDs a good candidate for the generation of entangled photon pairs. It is necessary that these QDs exhibit a pure radiative decay and single photon emission. We investigated both properties under pulsed excitation in a time-resolved μ-PL setup. Figure 4a depicts the PL decay curves obtained at 4K from the X line of a single dot emitting at 1.2778 eV for two different excitation powers. By fitting with a single exponential we extract a lifetime of ≈2ns, independently of the pump power as expected for single QDs. These lifetimes are comparable, within a factor of 2, to the lifetimes of standard (100) InAs QDs [28] emitting in the same energy range, demonstrating the high oscillator strength and quantum efficiency of these dots at 4K. For pump powers larger than ≈50μW, a prolonged rise-time is observed in the PL emission, as a consequence of state filling and subsequent cascade from higher excited dot levels [29]. The same effects are responsible for the strong saturation of the PL intensity, depicted in the inset of Fig.4a. Note that both the lifetime of 2ns and the demonstration of state filling is an indirect proof of the good optical quality of the (211) InAs/GaAs QDs, equivalent to their well known (100) InAs/GaAs counterparts.

Finally, single photon emission is a direct proof of the single exciton nature of investigated QD. The experiment was performed on a typical Hanbury-Brown-Twiss setup with two avalanche photodiodes (APD EG&G Model SPCM-AQR 13) [30]. Each APD is placed after a monochromator, spectrally filtering the exciton line and avoiding optical cross-talk [31]. The single dot emitting at 1.2778 eV is pumped with a power of 27μW, i.e. just below saturation, in order to maximize the photon count rate. The count rates were 6500 and 9000 counts per second (cps) on each APD with a background count rate of 2500 and 4000 cps, respectively, measured next to the exciton emission line. The integration time was 3600sec. Figure 4b shows the raw second order autocorrelation function $g^2(\tau)$ of the exciton line. At each repetition of the laser pulse we observe a correlation peak, the area under which normalizes to A(τ)=1±0.05 [32]. At τ=0, we observe an absence of coincidences, signature of the single photon emission from the QD under investigation. The normalized area of the τ=0 peak equals to A(τ=0)=0.28±0.05, which is less than 0.5, proof of single photon emission. Note that the $g^2(\tau)$ function has not been corrected for any background, and the relatively poor value of A(τ=0) is mostly due to the excess of background noise. This noise is under investigation, but may possibly arise from a heavily doped buffer layer.

In conclusion, we have shown that (211) InAs/GaAs quantum dots are a good candidate for the generation of entangled photon pairs. We have demonstrated that the mean FSS splitting of such quantum dots is below 10 μeV. These quantum dots have similar lifetimes with well known (100) quantum dots, and we have demonstrated single photon emission.

**Acknowledgements** This work was co–funded by the European Social Fund and National resources through the program HERAKLITUS I, and the C'NANO "Au fil de l'eau" program.

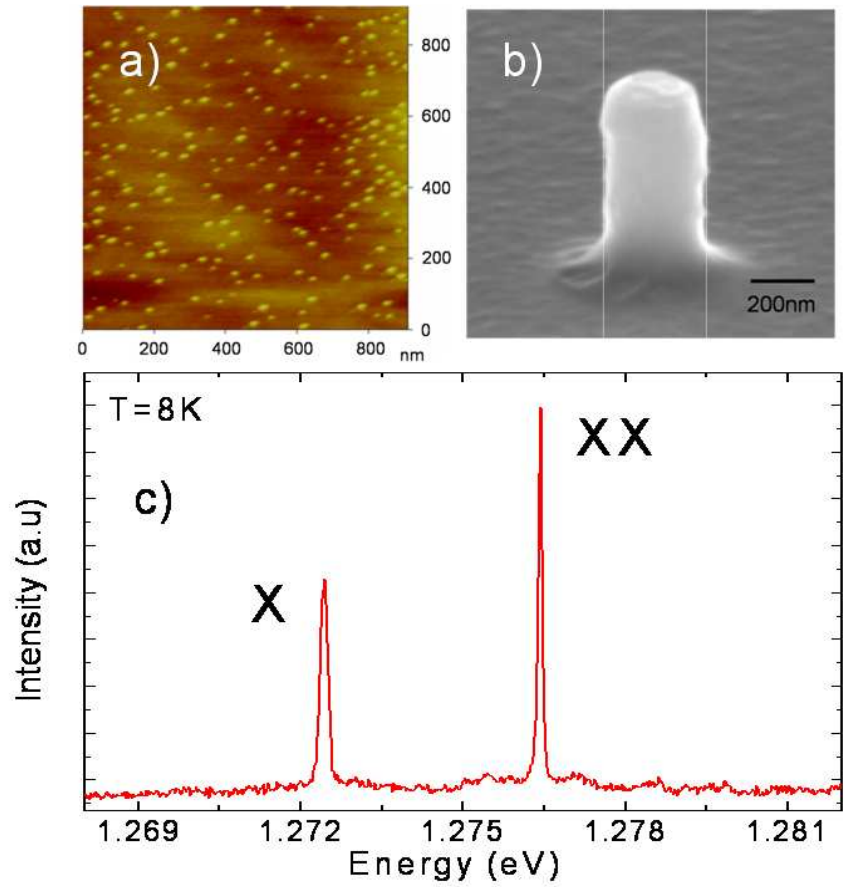

Fig.1

**Fig. 1.** (Color online) **a:** AFM image from an uncapped QD layer grown at the same conditions as in the sample used in this work. **b:** SEM picture from a typical mesa used for the μ-PL experiments. **c:** μ-PL spectrum from a single (211)B InAs/GaAs QD showing emission from excitons (X) and bi-excitons (XX).

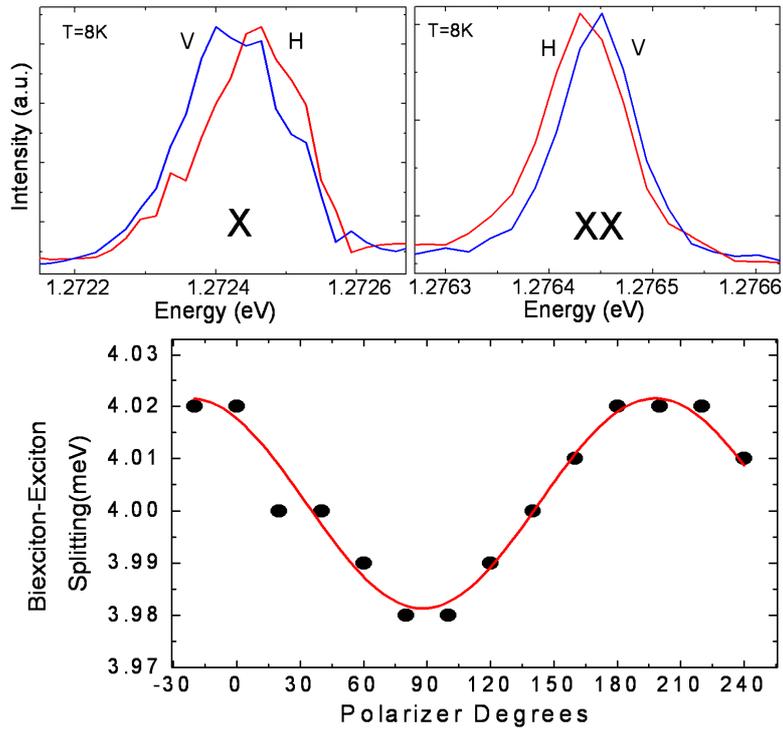

Fig.2

**Fig. 2.** (Color online) *Top panel*: Polarization-resolved emission from the exciton (X) and bi-exciton (XX) single dot lines of Fig.1c. *Bottom panel*: XX-X energy splitting as a function of polarization angle, showing characteristic periodicity. The extracted FSS for this QD is 20μeV.

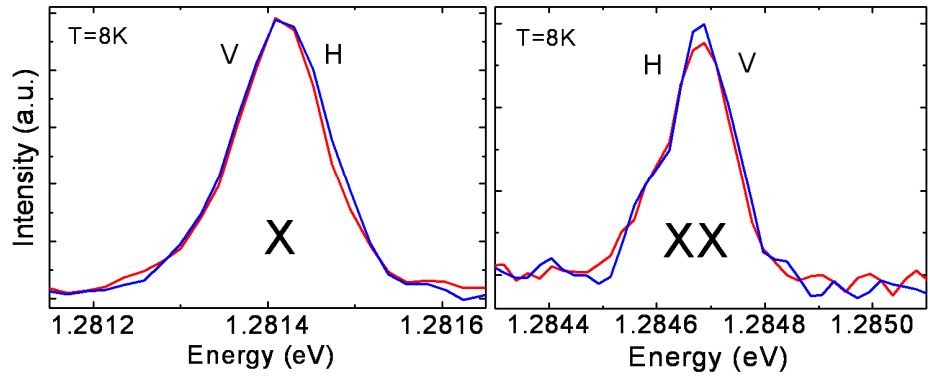

Fig.3

**Fig. 3.** (Color online) Polarization-resolved emission from the X and XX lines of a different (211)B InAs/GaAs QD exhibiting no measurable polarization dependence (FSS<10μeV).

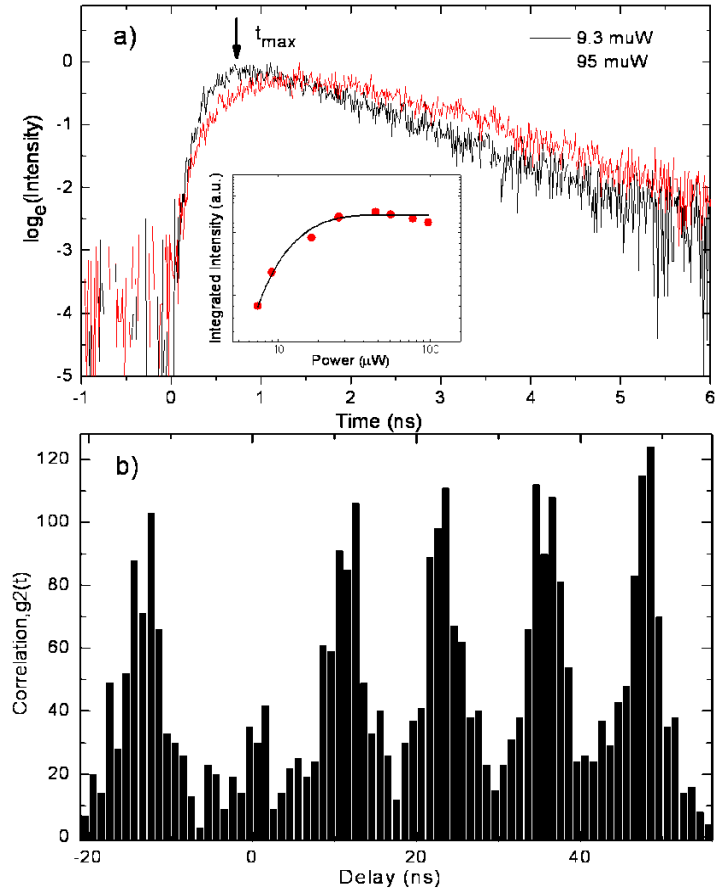

Fig.4

**Fig. 4.** (Color online) **a:** Time-decay curves obtained at T=4K from the exciton emission line of a single (211)B InAs/GaAs QD as a function of excitation power. The inset shows the time-integrated exciton PL intensity versus power, marking a clear saturation regime in this power range. **b:** Second-order correlation function $g^2(\tau)$ recorded at T=8K with 27μW pump power demonstrating a clear anti-bunching behavior characteristic of single dot emission.